
%
%
\input harvmac
%
%

\newif\ifIncludeFigs\IncludeFigsfalse
\newif\ifIncludeEPSF\IncludeEPSFfalse
%
%
%
\def\hasfigures{
   \def\yesans{y }
   \message{ This paper has associated figures. Include them (y/n)?}
   \read-1 to\answerrr
   \ifx\answerrr\yesans
        \IncludeFigstrue
   \fi
}
\def\haspsfigures{
   \def\yesans{y }
   \message{ This paper comes with encapsulated postscript figures.}
   \message{ Should they be included (requires epsf.tex and Rockicki
dvips)(y/n)?}
   \read-1 to\answerrr
   \ifx\answerrr\yesans
        \openin1=epsf.tex\relax \ifeof1
                \relax\message{ Could not find epsf.tex; no action taken.}
        \else
		\closein1
                \input epsf.tex
                \IncludeFigstrue
                \IncludeEPSFtrue
        \fi
   \fi
}
%
%

%
%

\def\iFig#1{\noindent\Fig#1 \hbox{: } {\reflabeL{ #1}}}
\def\Fig{Fig.~\the\figno\nFig}
\def\nFig#1{\xdef#1{Fig.~\the\figno}%
\writedef{#1\leftbracket Fig.\noexpand~\the\figno}%
\ifnum\figno=1\immediate\openout\ffile=figs.tmp\fi\chardef\wfile=\ffile%
\immediate\write\ffile{\noexpand\medskip\noexpand\item{Fig.\ \the\figno. }
\reflabeL{#1\hskip.55in}\pctsign}\global\advance\figno by1\findarg}

%
%

\haspsfigures           
\def\VEV#1{\left\langle #1\right\rangle}
\def\sec{\ifmmode \,\, {\rm sec} \else sec \fi}
\def\eV {\ifmmode \,\, {\rm eV} \else eV \fi}
\def\keV{\ifmmode \,\, {\rm keV} \else keV \fi}
\def\MeV{\ifmmode \,\, {\rm MeV} \else MeV \fi}
\def\GeV{\ifmmode \,\, {\rm GeV} \else GeV \fi}
\def\TeV{\ifmmode \,\, {\rm TeV} \else TeV \fi}
\def\pbarn{\ifmmode \,\, {\rm pb} \else pb \fi}
\def\deg{\ifmmode ^\circ\, \else $^\circ\,$ \fi}
\def\Mx{{m_{\tilde\chi}}}
\def\Mq{m_q}
\def\Msq{m_{\tilde q}}
\def\ra{\rightarrow}
\def\fun#1#2{\lower3.6pt\vbox{\baselineskip0pt\lineskip.9pt
  \ialign{$\mathsurround=0pt#1\hfil##\hfil$\crcr#2\crcr\sim\crcr}}}
\def\la{\mathrel{\mathpalette\fun <}}

\def\order{{\cal O}}
\def\etal{{\it et al.}}
\def\neut{{\tilde\chi}}

\def\tanb{\tan\beta}
\def\Msf{ m_{\tilde f}}

\overfullrule=0pt


%
%
\xdef\prenum{\vbox{ \hbox{EFI-93-38} \hbox{MAD-PH-766}
\hbox{IASSNS-HEP-93/37}
\hbox{hep-ph/9306325}}}
\Title{\prenum}{Neutralino Annihilation into Gluons}
\vskip -.4in
\def\madadd{\it Physics Department, University of Wisconsin,
        Madison, WI 53706}
\def\gjadd{\it Enrico Fermi Institute and Dept. of Physics
           5640 S. Ellis Ave., Chicago, IL 60637}
\def\mkadd{\it School of Natural Sciences, Institute for Advanced Study,
        Princeton, NJ 08540}

\centerline{ Manuel Drees$^a$\footnote{$^\dagger$}{drees@wiscphen.bitnet.},
Gerard
Jungman$^{b}$\footnote{$^\ddagger$}{jungman@yukawa.uchicago.edu.  After
Sept. 1, 1993: Dept. of
Physics, Syracuse University, Syracuse, NY 13244.}, Marc
Kamionkowski$^c$\footnote{$^\sharp$}{kamion@guinness.ias.edu}, and Mihoko
M.
Nojiri$^a$\footnote{$^\natural$}{nojiri@phenou.physics.wisc.edu.
After July 1, 1993: KEK Theory Group, Oho, Tsukuba, Ibaraki 305,
Japan.}}

\def\addspace{\vskip 0.05in}
\vskip .13in
\centerline{ $^a$\madadd}\addspace
\centerline{ $^b$\gjadd}\addspace
\centerline{ $^c$\mkadd}\addspace
\vskip .2in

\noindent

We present a complete calculation of the cross section for
neutralino annihilation into the two-gluon final state.
This channel can be quite important
for the phenomenology of neutralino annihilation due to the
well-known helicity suppression of neutralino
annihilation into light quarks and leptons.  In addition, we
calculate the cross section for annihilation of neutralinos into
a gluon and quark-antiquark pair, and discuss QCD corrections
to the tree-level cross sections for neutralino annihilation
into quarks.  If the neutralino is lighter than the top quark,
the effect of these results on high-energy neutrino
signals from neutralino annihilation in the Sun and in the
Earth can be significant, especially if the neutralino is
primarily gaugino.  On the other hand, our results should have
little effect on calculations of the cosmological abundance of
neutralinos.  We also briefly discuss implications for
cosmic-ray antiprotons from neutralino annihilation in the
galactic halo.
\vskip 0.2in
\vfill
\Date{June 1993}


%
%
\lref\darkmatter{For recent reviews of dark matter and its detection,
        see {\it Proceedings of the ESO-CERN Topical Workshop on
        LEP and the Early Universe}, ed. J.~Ellis, P.~Salati, and
        P.~Shaver (CERN preprint TH.5709/90),
        V.~Trimble, \sl Ann.
        Rev. Astron. Astrophys. \bf 25\rm, 425 (1989); J.~R.~Primack,
       B.~Sadoulet, and D.~Seckel, \sl Ann. Rev. Nucl. Part. Sci.
       \bf B38\rm, 751 (1988);  \sl Dark Matter in the Universe,
       \rm eds.  J.~Kormendy and G.~Knapp (Reidel, Dordrecht, 1989).}
\lref\haberkane{H.~E.~Haber and G.~L.~Kane,
        \sl Phys. Rep. \bf 117\rm, 75 (1985).}
\lref\barbieri{R.~A.~Barbieri and G.~F.~Giudice, \sl Nucl. Phys. \bf B306\rm,
       63 (1988).}
\lref\ellishag{J.~Ellis, J.~S.~Hagelin, D.~V.~Nanopoulos, K.~A.~Olive,
       and M.~Srednicki, \sl Nucl. Phys. \bf B238\rm, 453 (1984).}
\lref\heavy{K.~Griest,
        M.~Kamionkowski, and M.~S.~Turner, \sl Phys. Rev. D \bf
        41\rm, 3565 (1990).}
\lref\dreesnojiri{M.~Drees and M. M. Nojiri, {\sl Phys. Rev. D}
{\bf 47}, 376 (1993).}
\lref\neutrinos{J.~Silk, K.~Olive, and M.~Srednicki, {\sl Phys.
        Rev. Lett.}
        {\bf 55}, 257 (1985); T.~Gaisser, G.~Steigman, and
        S.~Tilav, {\sl Phys. Rev. D} {\bf 34}, 2206 (1986);
        J.~Hagelin, K.~Ng, and K.~A.~Olive, {\sl Phys. Lett. B} {\bf
        180}, 375 (1987);  M.~Srednicki, K.~Olive, and J.~Silk,
        {\sl Nucl. Phys.} {\bf B279},
        804 (1987); K.~Ng, K.~A.~Olive, and M.~Srednicki, \sl Phys.
        Lett. B \bf 188\rm, 138 (1987); K.~A.~Olive and M.~Srednicki,
        \sl Phys. Lett. B \bf 205\rm, 553 (1988); L.~Krauss,
        M.~Srednicki, and F.~Wilczek,
        {\sl Phys. Rev. D} {\bf 33}, 2079 (1986); K.~Freese,
        {\sl Phys. Lett. B} {\bf 167}, 295 (1986); F.~Halzen,
        T.~Stelzer, and M.~Kamionkowski, {\sl Phys. Rev. D} {\bf
        45},4439 (1992); G.~F.~Giudice and E.~Roulet, \sl Nucl. Phys.
        \bf B316\rm, 429 (1989).; G.~Gelmini, P.~Gondolo, and
        E.~Roulet, \sl Nucl. Phys. \bf B351\rm, 623 (1991).}
\lref\griest{K.~Griest, \sl Phys. Rev. \bf D38\rm, 2357 (1988);
        FERMILAB-Pub-89/139-A (E); \sl Phys. Rev.
        Lett. \bf 61\rm, 666 (1988).}
\lref\OandS{K.~A.~Olive and M.~Srednicki,
        \sl Phys. Lett. \bf B230\rm, 78 (1989);
        K.~A.~Olive and M.~Srednicki, \sl Nucl. Phys. \bf
        B355\rm, 208 (1991); J.~McDonald, K.~A.~Olive, and
        M.~Srednicki, {\sl Phys. Lett. B.} {\bf 283}, 80 (1992).}
\lref\antip{J.~Silk and M.~Srednicki, {\sl Phys. Rev. Lett.}
        {\bf 53}, 624 (1984); J.~Ellis \etal, {\sl Phys. Lett. B}
        {\bf 214}, 403 (1989); F.~Stecker, S.~Rudaz, and
        T.~Walsh, {\sl Phys. Rev. Lett.} {\bf 55}, 2622 (1985);
        F.~Stecker and A.~Tylka, {\sl Astrophys.
        J.} {\bf 336}, L51 (1989); S.~Rudaz and F.~Stecker, {\sl
	Astrophys. J.}
        {\bf 325}, 16 (1988); F.~Stecker, {\sl Phys. Lett. B} {\bf 201},
        529 (1988); M.~Srednicki, S.~Theissen, and J.~Silk,
        {\sl Phys. Rev. Lett.} {\bf 56}, 263 (1986); S.~Rudaz,
        {\sl Phys. Rev. Lett.} {\bf 56}, 2128 (1986);
        L.~Bergstrom and H.~Snellman, {\sl Phys. Rev. D} {\bf 37}, 3737
        (1988); L.~Bergstrom, {\sl Nucl. Phys.} {\bf B325}, 647 (1989);
        \sl Phys. Lett. B \bf
        225\rm, 372 (1989); G.F.~Giudice and K.~Griest, {\sl Phys. Rev. D}
        {\bf 40}, 2549 (1989); S.~Rudaz, {\sl Phys. Rev. D} {\bf 39},
        3549 (1989); A.~Bouquet, P.~Salati, and J.~Silk, {\sl Phys.
        Rev. D} {\bf 40}, 3168 (1989); M.~S.~Turner and F.~Wilczek,
        {\sl Phys. Rev. D} {\bf 42}, 1001 (1990); A.~J.~Tylka,
        {\sl Phys. Rev. Lett.} {\bf 63}, 840 (1989);
        M.~Kamionkowski and M.~S.~Turner, \sl Phys. Rev. D \bf
        43\rm, 1774 (1991).}
\lref\press{W.~H.~Press and D.~N.~Spergel, \sl Astrophys. J. \bf
        296\rm, 679 (1985); A.~Gould, \sl Astrophys. J. \bf
        321\rm, 571 (1987); A.~Gould, \sl Astrophys. J. \bf 388\rm, 338
        (1991).}
\lref\ritz{S.~Ritz and D.~Seckel, \sl Nucl. Phys. \bf B304\rm,
        877 (1988).}
\lref\pbar{S. P. Ahlen \etal, {\sl Phys. Rev. D} {\bf 61}, 145
(1988).}
\lref\tarle{G. Tarle, private communication.}
\lref\higgs{J.~F.~Gunion, H.~E.~Haber, G.~Kane, and S.~Dawson,
        \it The Higgs   Hunter's Guide, \rm (Addison-Wesley,
        Redwood City, 1990).}
\lref\pdb{Particle Data Group, {\sl Phys. Rev. D} {\bf 45}, S1 (1990).}
\lref\marcneutrinos{ M.~Kamionkowski, {\sl Phys.~Rev. D} {\bf 44}, 3021
       (1991).}
\lref\goldberg{H.~Goldberg, \sl Phys. Rev. Lett. \bf 50\rm, 1419
       (1983).}
\lref\experiments{For reviews of current and future
       energetic-neutrino detectors, see, e.g., {\it High Energy
       Neutrino Astrophysics,} edited by V.~J.~Stenger,
       J.~G.~Learned, S.~Pakvasa, and X.~Tata (World Scientific,
       Singapore, 1992).}
\lref\rudaz{J.~Ellis and S.~Rudaz, {\sl Phys. Lett.} {\bf 128B}, 248
            (1983).}
\lref\runmass{E.~ Braaten and J.~ Leveille, {\sl Phys. Rev.} {\bf D22},
               715 (1980); N. Sakai, {\sl Phys. Rev.} {\bf D22}, 2220
               (1980); T. Inami and T. Kubota,{\sl Nucl. Phys.}
               {\bf B179}, 171 (1981); S. Gorishnii, A.L. Kataev and
               S.A. Larin, {\sl Yad. Fiz.} {\bf 40}, 517 (1984);
               M. Drees and K.--I. Hikasa, {\sl Phys. Lett.}
               {\bf B240}, 455 (1990).}
\lref\kuehn{K.G.~Chetyrkin and J.H.~K\"uhn, {\sl Phys. Lett}
             {\bf B248}, 359 (1990).}

%
%
\newsec{Introduction}

Luminous matter almost certainly does not
account for all matter in the Universe \darkmatter, and
this matter deficit inspires both particle-physics and
astrophysics speculation. Among the particle-physics solutions
discussed, perhaps the most attractive idea is that
stable weakly interacting massive particles (WIMPs)
may make up the dark matter. Currently, the most promising
candidate WIMP is the neutralino \haberkane, a linear combination
of the supersymmetric partners of the photon, $Z^0$, and Higgs
bosons.

It is well known that the cosmological abundance of a WIMP is
inversely proportional to its annihilation cross section.  In
the earliest calculations, annihilation of neutralinos into
light fermions was considered \ellishag\griest; subsequently,
annihilation into gauge and Higgs bosons was taken into
account \heavy\OandS\dreesnojiri.  At this point,
the cross sections for annihilation into all two-body final
states which occur at tree level had been calculated.  The basic
conclusion of all these papers is that, in almost all
regions of supersymmetric parameter space, the neutralino
makes an excellent candidate for the dark matter in galactic
halos, and, in some cases, can provide an abundance suitable to
account for a flat Universe.

The neutralino annihilation cross section is also needed to
determine event rates in numerous schemes for indirect detection
of WIMPs in the galactic halo.  The existence of WIMPs in the
halo may be inferred through observation of distinctive
spectra of cosmic-ray antiprotons, $\gamma$-rays,
or positrons produced by
annihilation of WIMPs in the halo \antip.  Perhaps a more promising
method of detection involves observation of energetic-neutrinos
{}from WIMP annihilation in the Sun and/or in the
Earth \neutrinos\marcneutrinos.  If neutralinos reside in the
galactic halo, then they will be captured in the Sun and in the
Earth \press.  Annihilation of neutralinos therein would
produce, amongst other things, neutrinos whose energy is some
fraction of
the neutralino mass, which could be detected in current (e.g., IMB,
Kamiokande II, MACRO) or next-generation (e.g. DUMAND, AMANDA,
super-Kamiokande) experiments \experiments.

In this paper, we calculate the cross
section for annihilation of neutralinos into two gluons
($\neut\neut\ra gg$) and
discuss the implications for indirect-detection searches,
especially those involving observation of energetic-neutrinos
{}from neutralino annihilation in the Sun and in the Earth.  We
also calculate the cross section for annihilation into the
quark-antiquark-gluon final state ($\neut\neut\ra q\bar q g$), and
we improve the tree-level cross sections
for annihilation of neutralinos into quarks by including
leading-log QCD corrections.

There are a few simple reasons to believe that the two-gluon
final state may be important for indirect-detection experiments,
although we do not expect this final state to have much impact on
cosmological relic-abundance
calculations.  Neutralinos in the halo, Sun, and Earth move with
velocities $v \,\roughly{<}\, 10^{-3}$, so annihilation through $P$ waves
(i.e., angular momentum $l=1$) is suppressed by roughly
$v^2 \,\roughly{<}\, 10^{-6}$; in other words, annihilation can occur only
through an $S$ wave.  Neutralinos are majorana particles, so the
$S$-wave cross section for annihilation into light quarks and leptons is
suppressed by $(m_f/\Mx)^2$ \goldberg, where $m_f$ is the quark
or lepton mass and $\Mx$ is the neutralino mass.  On the other
hand, there is no such suppression of $S$-wave annihilation into
gluons.  Therefore, even though annihilation into gluons is
formally suppressed relative to that into quarks by
$\alpha_s^2$, the square of the strong coupling constant, in
practice, $S$-wave annihilation of neutralinos into gluons may
be comparable to, or even stronger than annihilation into quarks,
as first noted by Rudaz and Bergstrom \antip.  For given SUSY
parameters, if the neutralino annihilates primarily into light
fermions at tree-level,
we expect our results to be important \marcneutrinos.
This will be the case if the
neutralino is lighter than the $W$ boson.  It may also be the case
if the neutralino is primarily gaugino and heavier than the $W$
boson but lighter than the top quark \heavy.  In addition,
since $S$-wave annihilation into light fermions
is additionally suppressed in the limit of large neutralino mass
the $gg$ annihilation
channel should become increasingly important as the neutralino
mass is increased.

When neutralinos freeze out in the early Universe, they are moving
with velocities of order 0.5, so the cross section for annihilation
contains significant $P$-wave as well as $S$-wave contributions.
Since $P$-wave annihilation into light fermions is {\it not}
suppressed in this case, annihilation into gluons is suppressed by
$\alpha_s^2$ relative to that into light fermions.  Therefore,
inclusion of the $gg$ final state should have no more than a small
effect on the cosmological abundance, and thus we
do not consider these abundance calculations further.

In the following Section, we describe the calculation and
present results of the cross sections for $\neut\neut\ra gg$ and
$\neut\neut\ra q\bar q g$.  We also discuss the leading-log QCD
correction to the annihilation cross section for
$\neut\neut\ra q\bar q$.  We present our results for a neutralino
of arbitrary mass and mixing and for arbitrary squark masses and
mixings.  From these, the cross sections for any given
supersymmetric parameters can be obtained.
In Section 3,  we illustrate our results for a few simple
examples and discuss the importance of the new channels
in various regions of parameter space.  In Section
4, we review production of
energetic neutrinos from annihilation of neutralinos in the
Earth, and we discuss the effect of the new cross sections on
the predicted event rates.
In the final Section, we briefly summarize and comment on the
implications of our results for predicted fluxes of cosmic-ray
antiprotons produced by annihilation of neutralinos in the halo.

\newsec{Annihilation Cross Sections}

In this Section, we review some relevant supersymmetry (SUSY)
formalism and present results for the cross sections.
We use the conventions and notation of Ref.~\dreesnojiri.  There
are four
neutralinos which are linear combinations of $\tilde B$,
$\tilde W_3$, $\tilde h_1^0$, and $\tilde h_2^0$, the
supersymmetric partners of the $U(1)$ gauge field, the third
component of the $SU(2)$ gauge field, and Higgs fields,
respectively. In the $(\tilde B, \tilde W_3, \tilde h_1^0,\tilde
h_2^0)$ basis, the neutralino mass matrix is
\eqn\massmatrix{
\left(\matrix{M_1 & 0 & -m_Z s_W \cos\beta & m_Z s_W
\sin\beta\cr
0 & M_2 & m_Z c_W \cos\beta & -m_Z c_W
\sin\beta\cr
-m_Z s_W \cos\beta & m_Z c_W \cos\beta & 0 &
-\mu \cr
m_Z s_W\sin\beta & -m_Z c_W \sin\beta & -\mu &
0 \cr} \right),}
where $\mu$ is the higgsino mass parameter, $M_1$ and $M_2$ are
the gaugino mass parameters, $s_W=\sin\theta_W$, and
$c_W=\cos\theta_W$, and we adopt the
GUT relation, $M_2=(5/3) M_1 \tan^2\theta_W$.  The ratio of Higgs
vacuum expectation values is
$\tan\beta=\VEV{H_2^0}/\VEV{H_1^0}$.  The elements of the matrix
that diagonalizes the mass matrix are $N_{ij}$.  Unlike in
Ref.~\haberkane, we allow the neutralino mass eigenvalues to
take on negative values, so the $N_{ij}$ are always real.  We
denote the lightest neutralino by the subscript ``0'';
specifically, the lightest neutralino is
\eqn\lightestneutralino{
\neut=N_{01}\tilde B + N_{02}\tilde W_3 + N_{03}\tilde h_1^0 +
N_{04} \tilde h_2^0.}

\subsec{Cross section for $\neut\neut\ra gg$}

In this subsection we present the results of our calculation of the
cross-section for annihilation of non-relativistic neutralinos
into two gluons.  Our results are only for the limit of
vanishing relative incident velocity, $v\ra0$, and are therefore
suitable for use in indirect-detection calculations.  Although
they are not valid for relic-abundance calculations, they could
be used to estimate the magnitude of the annihilation
cross section at $v\sim0.5$.

\ifIncludeEPSF
{\topinsert
        \epsfysize=6in
        \centerline{\vbox{\epsfbox{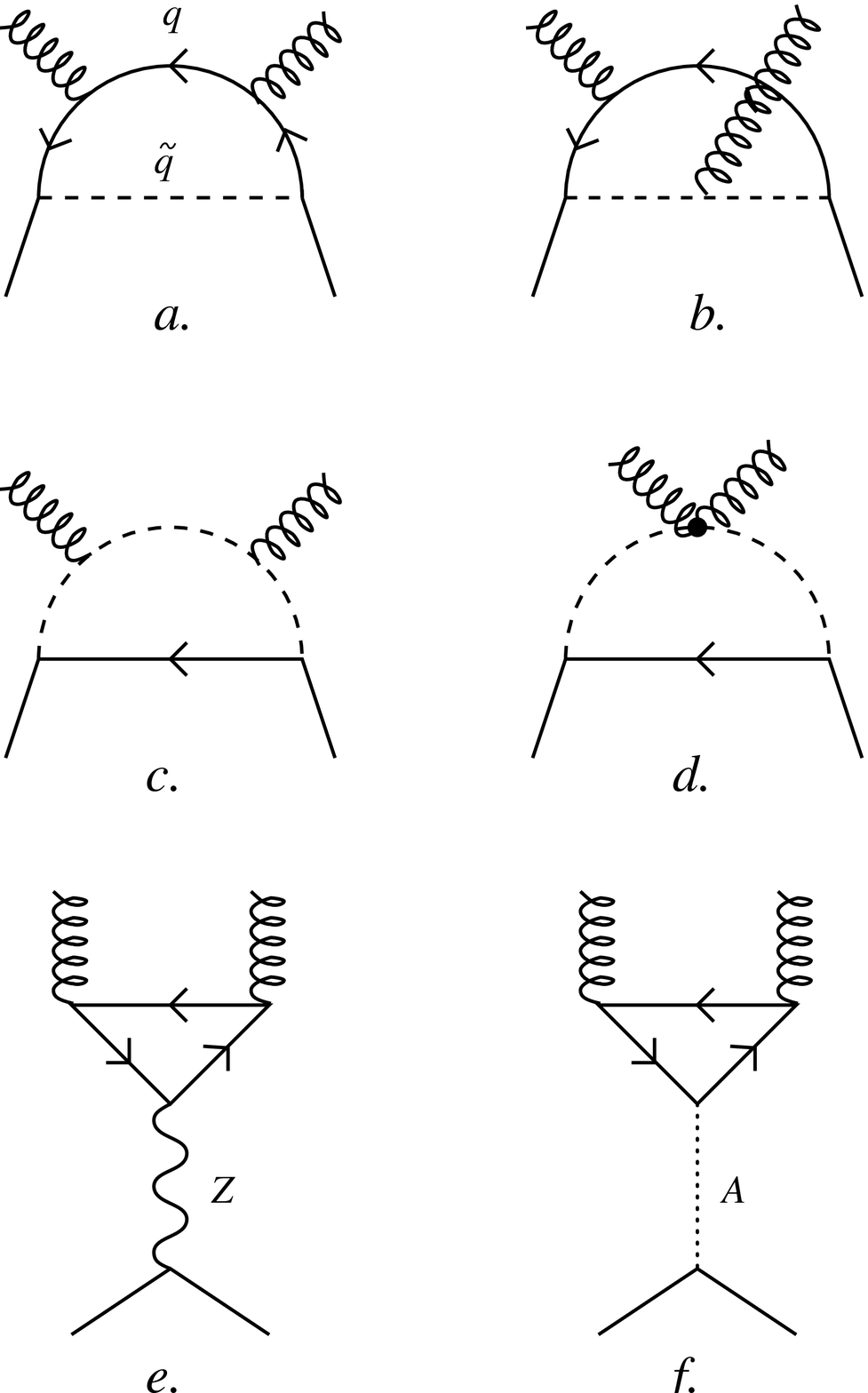}}}
        \vskip 0.2in
        \iFig\feynggfig{Feynman diagrams for
                annihilation of neutralinos into two gluons.
		Figures (a)-(d)
                show quark--squark loops, while (e) and (f) show
                the exchange of the $Z^0$ boson or the pseudoscalar
                Higgs boson $A^0$.}
        \epsfxsize=0pt
        \epsfysize=0pt
        \vskip .5cm
\endinsert
}
\else
        \nFig\feynggfig{Feynman diagrams for the one-loop
                annihilation of neutralinos into two gluons.
		Figures (a)-(d)
                show quark--squark loops, while (e) and (f)  show
                the exchange of the $Z^0$ boson or the pseudoscalar
                Higgs boson $A^0$.}
\fi

In \feynggfig, the one-loop diagrams for annihilation of neutralinos
into two gluons are displayed, and anti-symmetrization (symmetrization)
on the identical particles in the initial (final) state is understood.
Due to the helicity structure in the
$v\ra0$ limit, only diagrams (a), (b), (e), and (f)
contribute.  This is because the $CP$ eigenvalue of the initial
two-neutralino
state in this limit is $-1$.  The two gluons in the
final state could therefore be produced only by an operator of
the form
$G_{\mu\nu}^a \widetilde G^{\mu\nu b}$,
where $G_{\mu\nu}^a$ is the gluon field-strength
tensor.  Inspection of the Dirac structure of diagrams (c) and
(d) shows that they cannot give rise to such a term.
Furthermore, the Dirac structure of the
remaining diagrams also simplifies considerably in the $v\ra0$
limit. This allows us to write a relatively compact form for the
final result.


The matrix element for $\neut\neut\ra gg$ is
\eqn\matgg{
{\cal M}={1\over 4\pi^2}\epsilon(k_1,k_2,\epsilon_1,\epsilon_2)
(-1)^{{\bar h}+{1\over2}} \delta_{h,\bar h} {\tilde{\cal M}}
{g_s^2 \over 2} \delta_{ab}.
}
The imaginary part of ${\tilde{\cal M}}$ is
\eqn\matim{
\eqalign{
{\rm Im}\,{\tilde{\cal M}} = - \pi \sum_q&
\theta(\Mx^2- \Mq^2)\ln{1+\beta_q\over 1-\beta_q} \cr
& \times \Biggl\{
\sum_{\tilde q}\left[ {1\over \Msq^2 + \Mx^2 - m_q^2}
\left({1\over2} \right)\left(S_q
{\Mq^2\over \Mx^2} + D_q {\Mq
\over \Mx}\right) \right]\cr
& - {2\Mq\over\Mx} {g_{Aqq} g_{A\neut\neut} \over (4\Mx^2-m_A^2) } -
{\Mq^2 \over \Mx^2 m_Z^2} I_{3q} {g_{Z\neut\neut} g \over
\cos\theta_W} \Biggr\},\cr }
}
where $\beta_q = \sqrt{ 1-m_q^2/\Mx^2}$.
The real part of ${\tilde{\cal M}}$ is
\eqn\matre{
\eqalign{
{\rm Re}&\,{\tilde{\cal M}} = \sum_q \Biggl[ \sum_{\tilde
q}\Biggl\{ -{1\over
2\Mx^2} \int_0^1\,dx\Biggl( {S_q \over x} \ln \left|{x^2 a +x
(b-1-a)+1 \over -x^2 a + x(b-1+a)+1}\right|\cr
&+ {S_q b + D_q \sqrt{ab} \over 1+a-b}\left({1\over 1-x} + {1\over
1+x}\right) \ln \left |{x^2 a +x(b-a-1)+1 \over b+a (1-x^2)}\right|\cr
&+ {1\over 1-b+xa} \left[S_q b\left({1\over x} + {1\over
1-x}\right) + D_q \sqrt{ab}{1\over 1-x} \right] \cr
& \times \ln\left| { b \over x^2 a -x(a+b-1) -1} \right|\cr
&+{1\over b-1+ax}\left[S_q b \left( {1\over x} - {1\over
1+x}\right) - D_q\sqrt{ab} {1\over x+1}\right] \cr
& \times \ln\left| {b \over x^2a
+ x(b-1-a)+1} \right|\Biggr) \Biggr\}\cr
& + 2\, I\left({\Mq^2 \over \Mx^2}\right) {\Mq\over \Mx}
\left(2 {g_{Aqq} g_{A\neut\neut} \over (4\Mx^2 - m_A^2)} + {\Mq\over
\Mx} I_{3q} {g_{Z\neut\neut} g\over m_Z^2\cos\theta_W} \right)
\Biggr].\cr}
}
Note that there is a sum over quarks, and for each quark, there
is an additional sum over the two squarks in the term containing the
parametric integral.  We have defined
$S_q=A_q^2+B_q^2$, and $D_q=A_q^2-B_q^2$, and $A_q$ and $B_q$
are quark-squark-neutralino couplings defined below.
We have also defined $a\equiv \Mx^2/\Msq^2$ and
$b\equiv\Mq^2/\Msq^2$, and it is important to note that the sign
of $\sqrt{ab}$ is the sign of $\Mx$.

Generally, left and right squarks ($\tilde q_L$ and
$\tilde q_R$) may mix \rudaz, and the squark mass eigenstates are then
(see, e.g., Ref.~\dreesnojiri),
\eqn\squarkmixing{
\eqalign{ \tilde q_1 &= \tilde q_L \cos\theta_q + \tilde q_R
\sin\theta_q, \cr
\tilde q_2 &= -\tilde q_L \sin\theta_q + \tilde q_R
\cos\theta_q, \cr}}
where $\theta_q$ are the squark mixing angles.  The
squark-quark-neutralino couplings for the lighter squark
eigenstates, denoted by the subscript ``1'', are
\eqn\ABcoup{
\eqalign{A_q &= {1\over2} [\cos\theta_q (X_q + Z_q) +
\sin\theta_q (Y_q+Z_q)],\cr
B_q &= {1\over2} [\cos\theta_q(X_q - Z_q) +
\sin\theta_q (Z_q-Y_q)].\cr}}
The squark-quark-neutralino couplings for the heavier squark
eigenstate, denoted by a subscript ``2'', are obtained by making
the substitutions $\sin\theta_q\ra\cos\theta_q$ and
$\cos\theta_q\ra -\sin\theta_q$.  Here, we have defined
\eqn\XYcoup{
\eqalign{X_q &= -\sqrt{2}[g T_{3q} N_{02} - g' (T_{3q}-e_q)
N_{01}],\cr
Y_q &= \sqrt{2} g' N_{01} e_q.\cr
}}
For up-type quarks we have
\eqn\Zupcoup{
Z_{q}= - {g m_q N_{04} \over \sqrt{2} m_W \sin\beta},}
and for down-type quarks we have
\eqn\Zdowncoup{
Z_{q}= - {g m_q N_{03} \over \sqrt{2} m_W \cos\beta}.}
In the above, $T_{3q}=\pm1/2$ is the weak isospin of the quark, $e_q$ is
its electric charge, $g$ is the $SU(2)_L$ gauge coupling, $g'$ is
the $U(1)_Y$ gauge coupling, and $m_W$ is the $W$-boson mass.

The mass of the pseudoscalar Higgs boson, $A^0$ (sometimes
referred to in the literature as $H_3^0$), is $m_A$;
the couplings of the pseudoscalar Higgs boson to up-type quarks are
\eqn\gAqqup{
g_{Aqq} = - {g m_q \cot\beta \over 2 m_W},}
and the couplings to down-type quarks are obtained by making the
substitution $\cot\beta\ra\tan\beta$.  The coupling of the
pseudoscalar Higgs boson to the neutralinos is
\eqn\gAchichi{
g_{A\neut\neut} = {g\over2}(N_{02} - \tan\theta_W N_{01}) (N_{03}
\sin\beta - N_{04} \cos\beta),}
and the coupling of the $Z^0$ boson to the neutralinos is
\eqn\gZchichi{
g_{Z\neut\neut} = {g( N_{03}^2-N_{04}^2) \over 4 \cos\theta_W}.}

The function $I(x)$, which arises from the three-point
function in diagrams (e) and (f) of \feynggfig, is given by
\eqn\manuelI{
I(x)=\cases{ -{1\over 4} \left[ \ln^2 {1+\beta(x) \over
1-\beta(x)} - \pi^2 \right] & for $x\leq 1$,\cr
\left[\arctan{1\over\sqrt{x-1} } \right]^2& for $x>1$,}
}
where $\beta(x)=\sqrt{1-x}$.

Note that the integration over the Feynman parameter $x$ can
be transformed in many ways. In particular, by judicious
use of the transformation $x\ra 1-x$ in some of the terms
of the integrand, it is possible to improve the convergence
of numerical integration. A poor transformation of this type
would yield singularities at the endpoints which would, in
principle, cancel when integrated. As we have written it above,
the integrand has been transformed so that it has no singularities
at the endpoints, and this is the form most suitable for
numerical integration.

Given the matrix element above, the cross section times relative
velocity is
\eqn\ggcross{
\sigma_{gg} v = |{\tilde{\cal M}}|^2 { \alpha_s^2 \Mx^2 \over 8\pi^3}.
}

Although the calculation is lengthy and complicated, our result
for the cross section was obtained independently by
several of the authors.  In addition,
the diagrams for neutralino annihilation into two gluons are
similar to some that appear in the calculation for annihilation
into two photons.  Rudaz performed the
calculation for pure photinos and higgsinos in the limit of
large squark masses, Giudice and Griest obtained similar
results and generalized to neutralinos of arbitrary mixing, and
Bergstrom  performed the calculation for photinos
for arbitrary squark and photino masses \antip.
For large squark masses ($a,b\ll1$), the expression for the
real part of the matrix element, Eq.~\matre, simplifies to
\eqn\simplematre{
        \eqalign{
        {\rm Re}\, \tilde{\cal M} \simeq & \sum_q
                \Biggl\{ -{1\over \Msq^2}
                \sum_{\tilde q} \left[ -S_q + ({m_q^2 \over
                \Mx^2} S_q + {m_q \over \Mx } D_q)
                 I \left( {m_q^2 \over \Mx^2} \right)\right]\cr
        & + 2 I \left( {m_q^2 \over \Mx^2 } \right) \left[ {2
        m_q \over \Mx} {g_{Aqq} g_{A\neut\neut} \over
        (4 \Mx^2 - m_A^2) } + {m_q^2 \over \Mx^2 m_Z^2} I_{3q}
        g_{Z\neut\neut} {g \over \cos\theta_W} \right] \Biggr\}. }
}
If the neutralino is a pure photino, the terms due to $Z^0$
and $A^0$ exchange disappear, and $D_q\ra0$.
In this limit we reproduce Bergstrom's result \antip\ for the box
diagram numerically, but our expression \matre\ is significantly more
compact, involving only a single (rather than double) integral over
Feynman parameters. Note also that, even in the
limit $m_q\ra0$, the cross section is nonzero which reflects the
fact that there is no chirality suppression of $S$-wave neutralino
annihilation into gluons.

\subsec{Cross section for $\neut\neut\ra g q\bar q$}

In this subsection we give the cross-section for the
annihilation of neutralinos into a gluon and quark-antiquark
pair.  This cross section also remains finite as $m_q
\rightarrow 0$, as does the
gluon-pair cross section discussed in the previous subsection.
Moreover, it contains only a single strong-interaction vertex, i.e.
only one factor of $\alpha_s$.
If the neutralino is heavier than the top quark,
annihilation into top quarks is unsuppressed, and the $g \bar q
q$ final state will be negligible in comparison.  Therefore, we consider
only the case that the neutralino is lighter than the top quark,
and calculate this cross section in the limit of zero mass for the
final state quarks.

\ifIncludeEPSF
{\topinsert
        \epsfxsize=4in
        \centerline{\vbox{\epsfbox{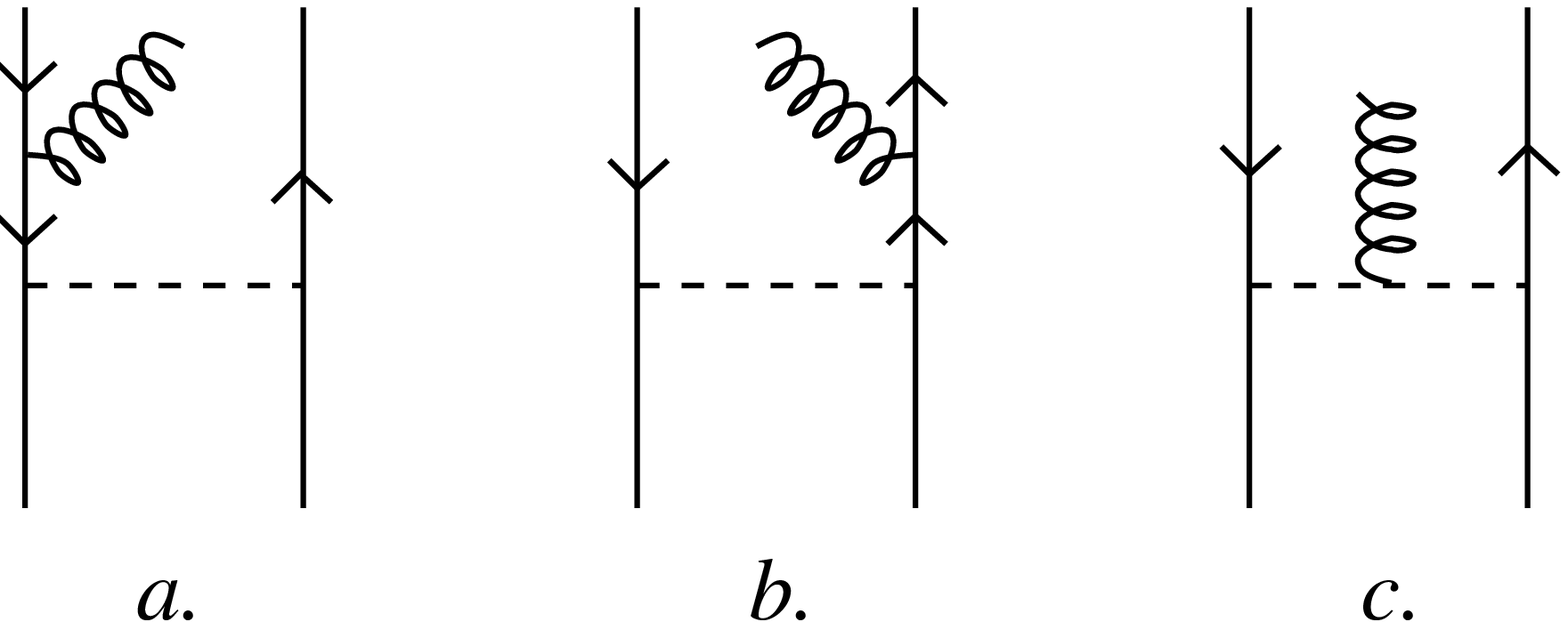}}}
        \vskip 0.2in
        \iFig\feyngqqfig{Feynman diagrams for the tree-level
                annihilation of neutralinos into a gluon and
                a quark antiquark pair.}
        \epsfxsize=0pt
        \epsfysize=0pt
        \vskip .5cm
\endinsert
}
\else
        \nFig\feyngqqfig{Feynman diagrams for the tree-level
                annihilation of neutralinos into a gluon and
                a quark antiquark pair.}
\fi

In \feyngqqfig, the diagrams for the cross section for
$\neut\neut\ra g \bar q q$ are displayed.  It is worth noting
that the expected infra-red divergence from soft-gluon emission
in these diagrams does not occur in the zero quark-mass limit.
Diagrams (a) and (b) do not vanish in this limit, but their
contributions are infra-red finite and combine
with the infra-red finite contribution of diagram (c).
As in \feynggfig, the anti-symmetrization on the identical
initial particles in \feyngqqfig\ is understood. Note that
the contributions from $Z^0$ or $A^0$ exchange in the
s-channel vanish in the massless-quark limit, so that the
diagrams of \feyngqqfig\ are the only ones to consider.

After some algebra, we reduce the calculation to the
following phase space integral;
\eqn\qqgcv{
        \eqalign{
        \sigma_{g\bar q q} v = \sum_{\tilde q}
        {8 A_q^4 \alpha_s \over  \pi^2 \Mx^2} &
        \int_0^1 dx_1\, \int_0^1 dx_2 \, \Theta(x_1+x_2-1) \cr
        &{(x_1+x_2 - 1)(x_1^2 + x_2^2 - 2(x_1 + x_2 - 1))
                \over (1-2x_1-r_{\tilde q})^2 (1-2x_2-r_{\tilde q})^2 },
        }
}
where $x_i \equiv E_i/\Mx$ are the normalized energies
of the quark and antiquark, $r_{\tilde q} \equiv \Msq^2 / \Mx^2$, and
$A_q$ is the quark-squark-neutralino coupling
defined above ($A_q=\pm B_q$ for massless quarks).
The displayed form of the integrand is most convenient as a
representation for the double differential cross-section,
$d\sigma / dx_1 dx_2$.
One of the remaining integrals in \qqgcv\ is elementary; integrating
this gives
\eqn\qqgcvtwo{
        \sigma_{g \bar q q} v = \sum_{\tilde q}
        {A_q^4 \alpha_s \over 2 \pi^2 \Mx^2}
        \int_0^1 dx\,
        { f_1(x) + f_2(x)\ln\bigl[ {1+r_{\tilde q} \over 1+r_{\tilde q}-2x }
                        \bigr]
                \over (1+r_{\tilde q}) (2x + r_{\tilde q} -1)^2 },
}
where
\eqn\qqgfs{
        \eqalign{
        f_1(x) &= -14x - 12 r_{\tilde q} x - 6 r_{\tilde q}^2 x
                + 18 x^2 + 2 r_{\tilde q} x^2 - 8 x^3, \cr
        f_2(x) &= 7 + 13r_{\tilde q} + 9 r_{\tilde q}^2
                + 3 r_{\tilde q}^3 - 12 x - 16 r_{\tilde q} x
                - 4 r_{\tilde q}^2 x + 4 x^2 + 4 r_{\tilde q} x^2.
        }
}
The remaining integral can be treated numerically in direct
fashion.

\subsec{Cross Section for $\neut\neut \ra f \bar f$}

For clarity and comparison, we present the cross section for
annihilation of neutralinos into fermions and
quarks\griest\heavy\dreesnojiri.  In the limit of zero relative
velocity, the cross section is
\eqn\ffcross{
\eqalign{
\sigma_{\bar f f} v = \sum_f & {c_f \theta(\Mx^2-m_f^2) \over 8
\pi} \Mx^2\sqrt{1-{m_f^2 \over \Mx^2 }} \cr
& \times \Biggl\{\sum_{\tilde f} \left[ {1 \over \Msf^2+\Mx^2
- m_f^2}\left(D_f + S_f {m_f \over \Mx} \right) \right ]\cr
& - {4
g_{A\neut\neut} g_{Aqq} \over 4\Mx^2 -m_A^2 } - 2
g_{Z\neut\neut} I_{3f} {m_q \over \Mx m_Z^2}{g \over
\cos\theta_W} \Biggr\}^2, \cr}}
where the sum on $f$ is over leptons as well as quarks, and the
color factor $c_f$ is 3 for quarks and 1 for leptons. This
cross section is proportional to the square of the quark mass.
(Recall that $|A_f| = |B_f|$ for massless fermions, i.e.
$D_f \propto m_f.$)
Also note that the cross section is proportional to the square of the
imaginary part of the amplitude for neutralino annihilation into
gluons, as it should be.

Previously, the cross section for neutralino annihilation into
quarks has been evaluated only at tree-level.  In addition to the tree-level
contributions,
there will be QCD corrections to this cross section which can be
important.
These QCD corrections can broadly be grouped into three classes:
Emission of a hard gluon at large angle to both quark and anti-quark;
emission of a hard gluon (almost) collinear to either quark or
anti-quark; and virtual corrections plus soft-gluon emission (these two
must be added to cancel infrared divergences). The first correction
has been treated in the previous subsection, where we saw that it
remains finite as $m_q \rightarrow 0$. The main effect of the second
contribution is to change the energy spectrum of the produced quarks;
this will be included using Monte Carlo results, as explained in
Sec. 4. Finally, the third class of corrections only renormalizes
the overall value of the cross section. As usual, the leading
logarithms in this correction can be most easily taken into account
by introducing ``running parameters''. This procedure, which is based
on the renormalization group, also automatically re-sums all powers
of these logarithms as they arise in higher orders of perturbation
theory.

In the case at hand there are two parameters that ``run'' due to
QCD corrections: The Yukawa contribution $Z_q$ to the LSP-quark-squark
couplings, and the quark mass $m_q$. Eqs. (2.9), (2.10) show that
Yukawa couplings and masses are proportional, and thus
run in the same way. Running quark masses have previously been
utilized in calculations of Higgs boson decay widths \runmass, but
they are also useful for the treatment of QCD corrections to the
production of massive quark pairs in $e^+ e^-$ annihilation \kuehn,
which more closely resembles our case of LSP annihilation.

For momentum
transfer $m_b<Q<m_t$, where $m_b$ and $m_t$ are the bottom- and
top-quark masses, the running masses are given by \runmass
\eqn\runningbmass{
m_b(Q)=m_b(m_b) \left( {\alpha_s(Q) \over \alpha_s(m_b)}
\right)^{12/23},}
and
\eqn\runningcmass{
m_c(Q)=m_c(m_c) \left( {\alpha_s(m_b) \over \alpha_s(m_c)}
\right)^{12/25} \left( {\alpha_s(Q) \over \alpha_s(m_b)} \right)^{12/23};}
if $Q>m_t$, the last factor becomes
\eqn\lastfactor{
\left( {\alpha_s(Q) \over \alpha_s(m_b) } \right)^{12/23} \ra \left( {
\alpha_s(m_t) \over \alpha_s(m_b) } \right)^{12/23} \left( {
\alpha_s(Q) \over \alpha_s(m_t) } \right)^{12/21},}
where $m_b(m_b)=4.5$ GeV and $m_c(m_c)=1.35$ GeV are the on-shell
quark masses, and $\alpha_s(Q)$ is the strong coupling constant
at scale $Q$ \pdb. We use $Q = \Mx$ in our calculation.

For neutralino masses in the range of $\order(10-100)$ GeV, the
running of the quark masses is significant, and since the
tree-level cross section is proportional to the square of the
quark mass, this affects the cross sections for annihilation
into quarks significantly.  For example, if the neutralino mass
is 80 GeV, the running $c$-quark mass is about 3/5 its tree-level
value, and the running $b$-quark mass is about 3/4 its
tree-level value, so the cross section for annihilation into
quarks is roughly half
that suggested by using the tree-level masses.  The consequences
for the rate of energetic-neutrino production from neutralino annihilation
will be discussed below.  The running of the quark masses should
have little effect on cosmological relic-neutralino abundances
because the cross section for neutralino annihilation through a
$P$ wave is not governed by the quark mass.

\newsec{Illustrative Examples}

The cross sections for the various annihilation
channels are complicated and depend on many parameters
making it difficult to analytically assess the relative
importance of each channel for arbitrary input parameters.
Therefore, to illustrate the possible importance of the new
channels, at least for a neutralino of some given simple composition,
we consider several specific examples.

The first limit we consider is that where the neutralino is a
pure photino:
$N_{01}=\cos\theta_W$, $N_{02}= \sin\theta_W$, and
$N_{03}=N_{04}=0$.
In addition, we also assume that the light-fermion masses are
small ($m_q\ll\Mx$), and that the photino mass is negligible compared
with the squark and top-quark masses ($\Mx\ll\Msq,m_t$).
Furthermore, the pure photino has no coupling to the $Z^0$ or $A^0$
bosons.
We also take degenerate squarks and assume no squark mixing.
Then, $D_q=0$ and $S_q=4\pi \alpha e_q^2$,
where
$\alpha$ is the electromagnetic fine-structure constant.
We also
note that $I(x)\simeq 1/x$ for $x\gg1$.  With these
assumptions, the cross section for photino annihilation into
gluons is
\eqn\sigmaggforphotino{
  \sigma_{gg}v \simeq 4 \alpha^2 \alpha_s^2 {\Mx^2
  \over \Msq^4}.}
With the same assumptions, annihilation into fermions occurs
primarily into $\bar\tau\tau$ pairs and, to a smaller degree,
into  $\bar c c$, and $\bar b b$ pairs.
The cross section for annihilation into these channels is
\eqn\sigmaffforphotino{
        \eqalign{
        \sigma_{\bar f f} v \simeq &\sum_f {8\pi c_f \alpha^2 e_f^4
                m_f^2   \over  \Msf^4,}\cr
        \simeq & {120\,\alpha^2\over\Msf^2}
                \left({ 1 \GeV \over \Msf }\right)^2.
        }
}
We have not used running quark masses here.  Use of running quark masses
decreases this estimate only slightly due to the slow running
of the $\tau$-lepton mass.  With similar
assumptions, the cross section for annihilation into the 3-body
final states are suppressed by larger inverse powers of the
squark mass.

Comparing the annihilation cross sections for the pure photino,
we find the following ratio:
\eqn\photinoratio{
  {\sigma_{gg} \over \sigma_{\bar f f} } \simeq \left( {\Mx
  \over 45\, {\rm GeV} } \right)^2,}
valid for $m_b\ll \Mx\ll m_t$, and $\Mx\ll\Msq$.  Therefore, the
relative importance of the gluon annihilation channel increases
roughly with the square of the photino mass and becomes
dominant for photino masses greater than about 45 GeV.

We now consider the pure $B$-ino limit: $N_{01}=1$ and
$N_{02}=N_{03}=N_{04} =0$.  If we use the GUT relation $M_1=(5/3)
M_2\tan^2\theta_W$ and if $M_2\ll \mu$, then the neutralino is
usually primarily $B$-ino.  In fact, in most cases where the
neutralino is primarily gaugino, it more closely resembles a
$B$-ino than a photino.  Once again, we take $m_b\ll\Mx\ll
m_t,\Msq$.  In this case, the cross section for $B$-ino
annihilation into gluons is
\eqn\sigmaggforbino{
  \sigma_{gg}v \simeq \left( {49\over 36} \right)^2 {g'^4 \over
  \Msq^4} {\alpha_s^2 \Mx^2 \over 8 \pi^3},}
while the cross section for annihilation into light fermions is
\eqn\sigmaffforbino{
  \sigma_{\bar f f}v \simeq {1\over 8\pi^3 \Msf^4} \sum_f c_f
  m_f^2 (S_f^L +S_f^R)^2 \simeq0.3\,{g'^4\over \Msf^2} \left( {
  {1 \GeV} \over \Msf} \right)^2.}
Again, using bare quark masses, we find the ratio
\eqn\binoratio{
  {\sigma_{gg} \over \sigma_{\bar f f}} \simeq \left( { \Mx
  \over 50\, {\rm GeV}} \right)^2,}
a result similar to that in the photino case.

The next limit we consider is the pure-higgsino limit:
$N_{01}=N_{02}=0$, and $N_{03}=N_{04}=1/\sqrt{2}$.  We make the
same assumptions about the masses.  Again, in this limit,
$D_q=0$.  The main contribution to the $gg$ cross section
comes from $t$-quark loops, unless $\tan \beta$ is very large, and the cross
section is roughly
\eqn\sigmaggforhiggsino{
  \sigma_{gg}v \simeq {\Mx^6 S_t^2 \alpha_s^2 \over 18 \pi^3 \Msq^4
  m_t^4},}
where we have assumed $\Mx^2 \ll m_t^2$.
For the light-fermion final states, annihilation
occurs primarily into $b$ quarks, and
\eqn\sigmaffforhiggsino{
  \sigma_{\bar f f} \simeq {3 S_b^2 m_b^2 \over 2 \pi \Msq^4}.}
Using on-shell quark masses we find that
\eqn\higgsinoratio{
  {\sigma_{gg} \over \sigma_{\bar f f}} \simeq \left( { \Mx
  \over 20\, {\rm GeV}} \right)^6 \cot^4 \beta.}
If we use running quark masses, the number is closer to 15 GeV.
We caution, however, that the LSP is only very rarely sufficiently
pure higgsino for Eq.~\higgsinoratio\ to be applicable even approximately.

In the previous examples, we have considered neutralinos
with no coupling to the $Z^0$ or $A^0$ bosons.  Now let us consider
what happens if the squarks are heavy enough that annihilation into
gluons and fermions occurs primarily through, for example,
the $A^0$-boson resonance.
Since two neutralinos in an $S$-wave cannot produce a physical
$Z^0$ boson, the contribution from $Z^0$ exchange is almost always
subdominant.
Again, we consider $m_b\ll \Mx\ll m_t$.
Then, decay of the intermediate virtual $A^0$ occurs primarily
through the top-quark loop.  Of the final-state light fermions,
decay occurs primarily into $\bar b b$ pairs.  With this
information, we find the ratio
\eqn\schannelratio{
  {\sigma_{gg} \over \sigma_{\bar b b}} \simeq {\alpha_s^2 \over
  3\pi^2 \tan^4\beta} \left( {\Mx \over m_b} \right)^2 \simeq
  {0.15 \over \tan^4 \beta} \, \left( {\Mx \over 100\, {\rm GeV}}
  \right)^2.}
Therefore, if the neutralino annihilates primarily through the
$A^0$, we do not expect the gluon final state to be significant.
Note that this can be a somewhat artificial example since it requires that
$2\Mx$ be near $m_A$ for the contribution from the intermediate
$A^0$ to be dominant.

In the limit of large squark masses, the cross section for
neutralino annihilation into the $\bar q q g$ 3-body final state
is suppressed relative to those for annihilation into the
light-quark and two-gluon final states by a factor of $\left(
{\Mx \over \Msq} \right)^4$.
Although the matrix elements from diagrams (a) and (b) in
\feyngqqfig\ are each inversely proportional to $\Msq^2$ in the
large squark-mass limit, these leading-order contributions
cancel and the contribution from diagrams (a) and (b)---as well
as that from diagram (c)---to the matrix element is proportional
to $\Msq^{-4}$ in the large squark-mass limit.  Therefore, if
the squark masses are large, the cross section for annihilation
into the 3-body channels is smaller than
those for annihilation into the $gg$ and $\bar q q$ channels.

\ifIncludeEPSF
{\topinsert
        \epsfxsize=4in
        \centerline{\vbox{\epsfbox{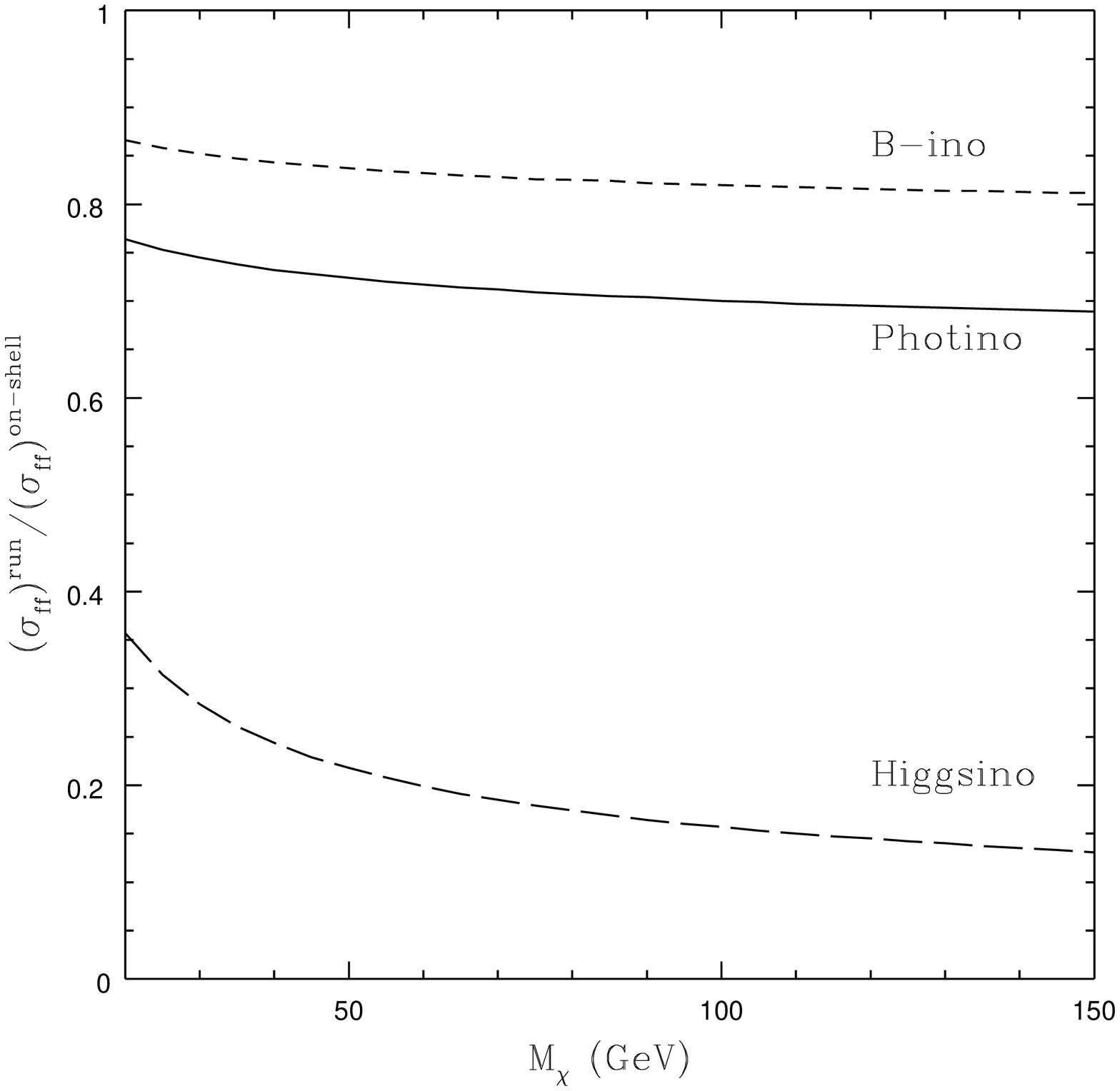}}}
        \vskip 0.2in
        \iFig\massplot{Effect of QCD corrections to the cross
                section for annihilation of higgsinos, photinos,
		and $B$-inos into light fermions ($\bar \tau
		\tau$, $\bar c c$, and $\bar b b$ pairs).}
        \epsfxsize=0pt
        \epsfysize=0pt
        \vskip .5cm
\endinsert
}
\else
        \nFig\massplot{Effect of QCD corrections to the cross
                section for annihilation of higgsinos, photinos,
		and $B$-inos into light fermions ($\bar \tau
		\tau$, $\bar c c$, and $\bar b b$ pairs).}
\fi

We now illustrate the effect of using running quark masses in
the cross section for annihilation of neutralinos into
$\bar\tau \tau$, $\bar c c$, and $\bar b b$ pairs.  If the
neutralino is a pure photino, then $\sigma_{\bar f f} \propto
c_f m_f^2 e_f^4$.  Quarks have fractional charge, so
annihilation occurs primarily into $\bar \tau \tau$ pairs, and
use of the running quark masses decreases the total cross section for
annihilation into light fermions only slightly.  The same is
true if the neutralino is pure $B$-ino.

On the other hand, if
the neutralino is a pure higgsino, the squark-quark-neutralino
coupling is a Yukawa coupling and is proportional to the quark
mass.  Therefore, the cross section for annihilation into $\bar
f f$ is proportional to $m_f^6$, so annihilation occurs
primarily into $\bar b b$ pairs.  If $m_b^{run}$ is the running
quark mass, and $m_b^{on-shell}$ is the on-shell quark mass, use of
running quark masses decreases the annihilation cross section by
a factor of $(m_q^{run}/m_q^{on-shell})^6$.  This results in a
dramatic decrease in the cross section for annihilation into
quark-antiquark pairs.  In practice, unless $M_2$ is extremely
large, a
neutralino that is primarily higgsino will contain some gaugino
component, so the actual decrease in the $\bar q q$ annihilation
cross section will hardly ever be as dramatic as indicated by the
pure-higgsino case.

In \massplot, we show the effect of QCD corrections to the cross
section for annihilation into $\bar \tau \tau$, $\bar c c$, and
$\bar b b$ pairs.  The ratio of the cross section using running
quark masses to that using on-shell quark masses is plotted against
neutralino masses from 20 to 150 GeV, and we illustrate the
results for the cases of a pure photino, $B$-ino, and higgsino.
The graph suggests that the effect of QCD corrections
to the tree-level cross sections is at least about 10\%, and
may, in some cases, be as big as 90\%.  In the next Section we
will discuss the effect of these results on the
energetic-neutrino signals.

\ifIncludeEPSF
{\topinsert
        \epsfxsize=4.5in
        \centerline{\vbox{\epsfbox{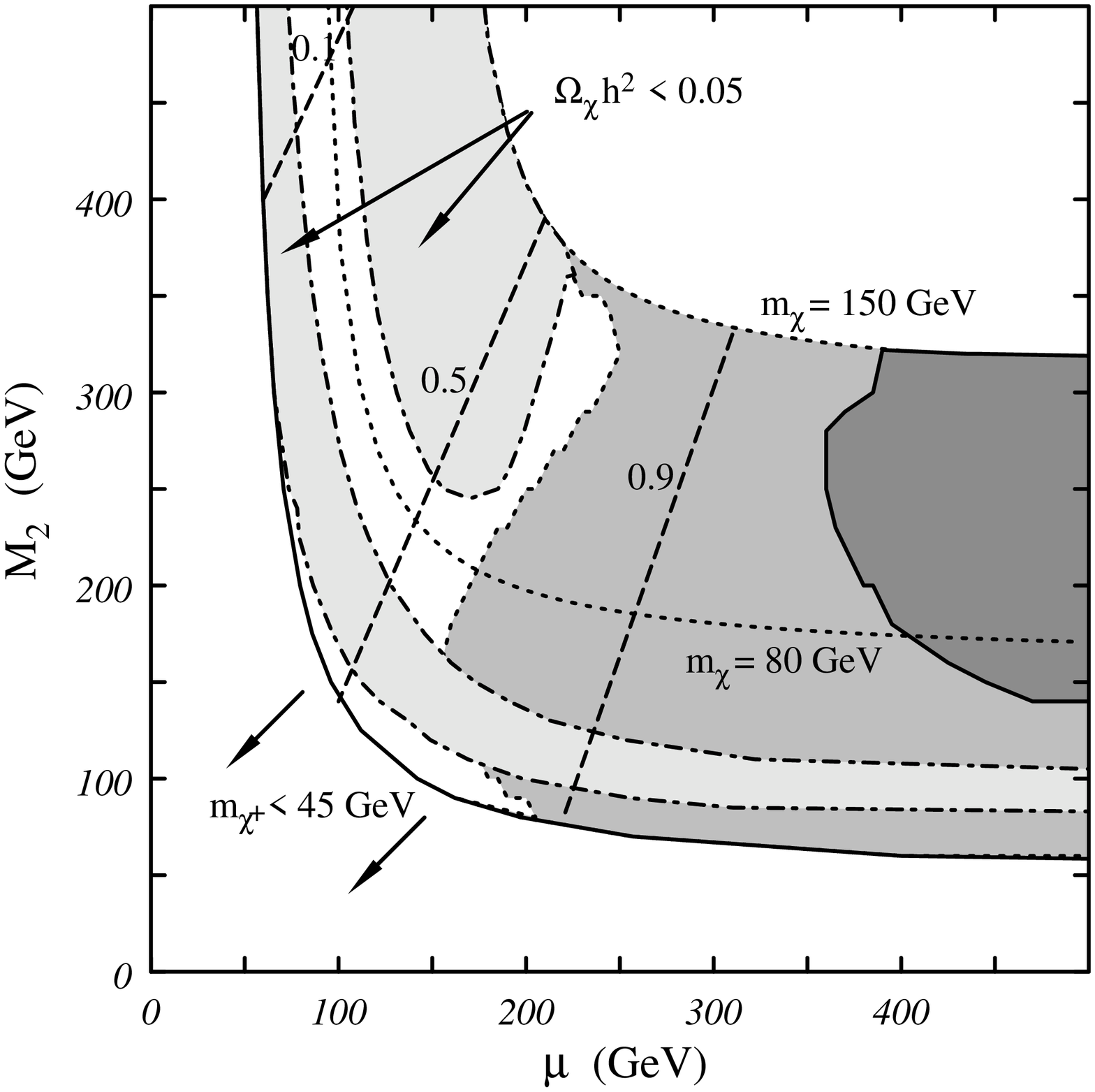}}}
        \vskip 0.2in
        \iFig\ggplot{Strength of $gg$ and $g \bar q q$ final
                states relative to the light-fermion final
		states. The darkest region corresponds to
		$\sigma_{gg}+\sigma_{g\bar q q} >\sigma_{\bar f f}$,
		the next lightest to
		$\sigma_{gg}+\sigma_{g\bar q q}>0.1\sigma_{\bar f f}$,
		and the lightest region corresponds to
		$\Omega_\chi h^2 < 0.05$, as indicated.
		Also shown are
		contours of $\Mx ={\rm 80, 150} \GeV $ and
		gaugino fraction contours at (0.9, 0.5, 0.1). See the
		text for discussion of other parameters. }
        \epsfxsize=0pt
        \epsfysize=0pt
        \vskip .5cm
\endinsert
}
\else
        \nFig\ggplot{Strength of $gg$ and $g \bar q q$ final
                states relative to the light-fermion final
		states. The darkest region corresponds to
		$\sigma_{gg}+\sigma_{g\bar q q}>\sigma_{\bar f f}$,
		the next lightest to
		$\sigma_{gg}+\sigma_{g\bar q q}>0.1\sigma_{\bar f f}$,
		and the lightest region corresponds to
		$\Omega_\chi h^2 < 0.05$, as indicated.
		Also shown are
		contours of $\Mx ={\rm 80, 150} \GeV $ and
		gaugino fraction contours at (0.9, 0.5, 0.1).
		See the
                text for discussion of other parameters.}
\fi

\ifIncludeEPSF
{\topinsert
        \epsfxsize=4.5in
        \centerline{\vbox{\epsfbox{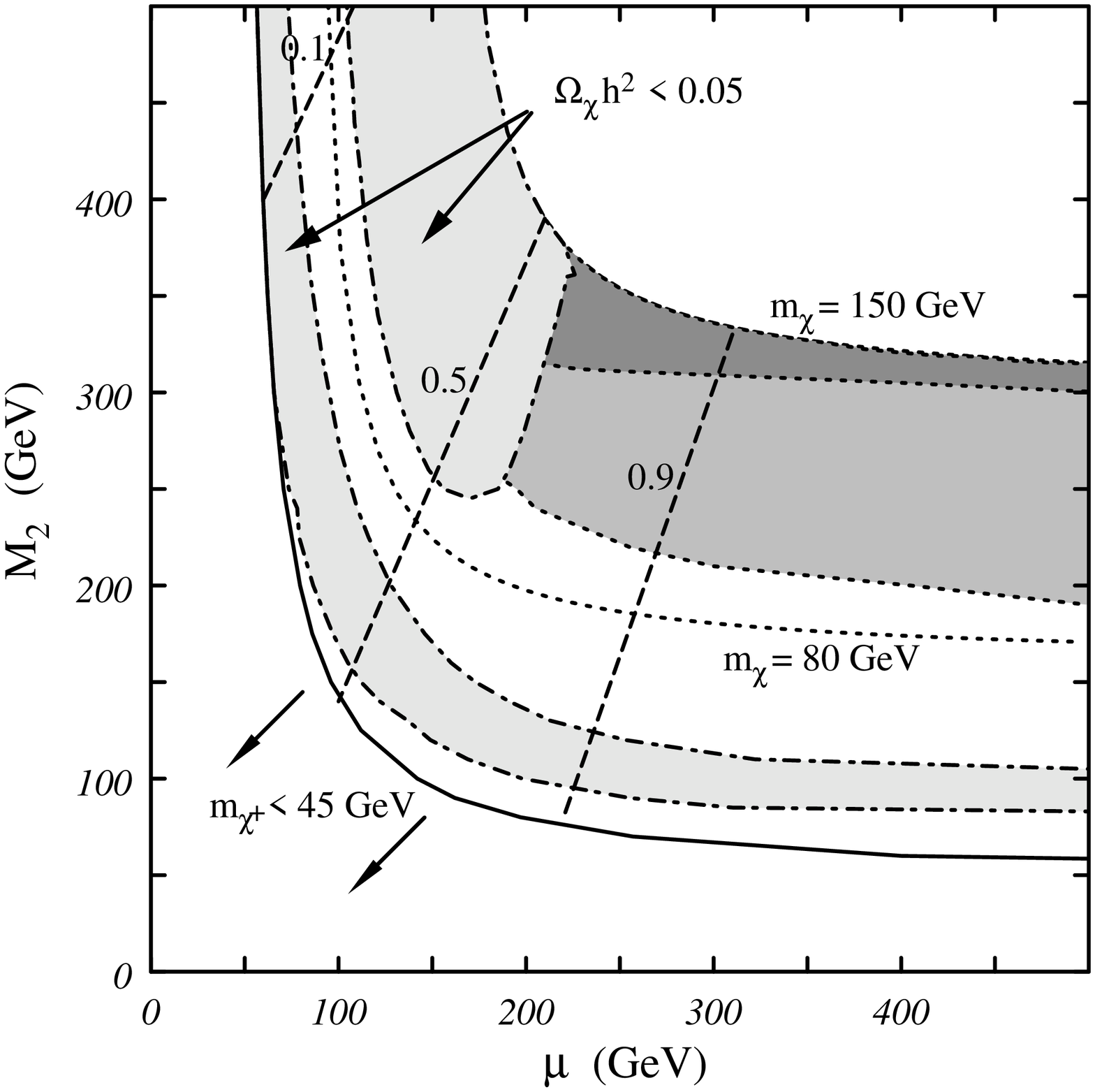}}}
        \vskip 0.2in
        \iFig\qqgplot{Strength of the $g \bar q q$ final
                state relative to the $gg$ final state.
		The darkest
		region corresponds to $\sigma_{qqg} / \sigma_{gg} > 1$,
		the next lightest to $\sigma_{qqg} / \sigma_{gg} > 0.1$,
		and the lightest to $\Omega_\chi h^2 < 0.05$, as in
		\ggplot.}
        \epsfxsize=0pt
        \epsfysize=0pt
        \vskip .5cm
\endinsert
}
\else
        \nFig\qqgplot{Strength of the $g \bar q q$ final
                state relative to the $gg$ final state.
		The darkest
		region corresponds to $\sigma_{qqg} / \sigma_{gg} > 1$,
		the next lightest to $\sigma_{qqg} / \sigma_{gg} > 0.1$,
		and the lightest to $\Omega_\chi h^2 < 0.05$, as in
		\ggplot.}
\fi

In \ggplot, we illustrate the importance of the $gg$ and $g \bar
q q $ annihilation channels relative to the light-fermion
annihilation channels for a neutralino of more arbitrary
composition.  In the regions shaded most heavily, the cross
section for neutralino annihilation into the $gg$ and $g\bar q
q$ final states is greater than that for annihilation into light
fermions, $\sigma_{gg}+\sigma_{g\bar q q}>\sigma_{\bar f f}$.  In the
more lightly shaded regions, the cross sections for the new
channels are at least 0.1 times as big as the cross section for
annihilation into light fermions, $\sigma_{gg}+\sigma_{g \bar q
q} > 0.1\,\sigma_{\bar f f}$.  In the most lightly shaded
regions, the cosmological relic abundance of the neutralino is
$\Omega_\neut h^2<0.05$, where $\Omega_\neut$ is the
cosmological mass density of neutralinos in units of the
critical density, and $h$ is the Hubble parameter in units of
100 km/sec/Mpc.  In these regions of parameter space, the
neutralino is viable, but its abundance is too small to account
for the dark matter in galactic halos.
In the empty region below and to the left of the
solid curve, the mass of the chargino is less than 45 GeV, so
this region is not experimentally viable.  In the empty region
in the upper right-hand corner of the graph, the neutralino mass
is greater than the top-quark mass (which we take to be 150
GeV), so our results will have little effect.  The short-dash
curve indicates the $\Mx=80$-GeV contour, and the diagonal
long-dash curves are of gaugino fractions ($\equiv N_{01}^2 +
N_{02}^2$) of 0.1, 0.5, and 0.9, as labelled.  Therefore, in the
bottom right part of the graph,
the neutralino is predominantly gaugino, and in the upper left
part, the neutralino is predominantly higgsino.  We took
$\tanb=2$, $m_A=500$ GeV, $m_t=150$ GeV, and $\Msq=200$ GeV.
We also took all squark and slepton masses to be degenerate and
assumed no mixing of right and left squarks, and we used running
quark masses to implement the leading-log QCD corrections.

Note that the new
channels can be quite important for a large range of masses and
neutralino mixings, and are especially important if the
neutralino is primarily gaugino.  Note that in large regions of
parameter space where the neutralino makes a good dark-matter
candidate, our results are significant.  We should also mention that
changes to \ggplot\ should be small if we change the assumed squark
mass.  This is because in the heavy-squark limit,
$\sigma_{gg}/\sigma_{\bar f f}$ is squark-mass independent.  In
addition, in the regions where we show $\Omega_\neut
h^2\la 0.05$, neutralino annihilation occurs primarily into
gauge bosons (where the neutralino is heavier than the $W$-boson
and primarily higgsino) or into light quarks through gauge and/or Higgs
bosons (in the regions where the neutralino mass is roughly 50
GeV).   The cross sections for these processes do not depend on
the squark mass.

In \qqgplot, we illustrate the importance of the 3-body final
state relative to the $gg$ final state for the same SUSY
parameters that were used in \ggplot.  The most heavily shaded
regions are those where the cross sections for annihilation
into $g\bar q q$ are larger than that for annihilation into
gluons, $\sigma_{g \bar q q} > \sigma_{gg}$.  In the more
lightly shaded regions, $\sigma_{g \bar q q}>0.1\,\sigma_{gg}$.
Again, the most lightly shaded regions are those where the relic
abundance of the neutralino is too small to account for the dark
matter in galactic halos.  Note that the 3-body channel seems to
be most important only in the regions where the
new channels do not have much effect (c.f., \ggplot).  If larger
squark masses were used, the relative importance of the 3-body
final state would be even smaller.  This suggests that
the effect of the 3-body final state is generally, although not
always, negligible.

We should point out that, in \ggplot\ and \qqgplot, we illustrated
the importance of the new annihilation channels under specific
assumptions about several SUSY parameters.  In general, there is
a large viable range for several of the parameters, and the
relative importance of the new channels for a given set of
assumptions may be either greater or smaller than
indicated in the limited regions of parameter space that we have
explored.

\newsec{Energetic Neutrinos From the Sun and the Earth}

If neutralinos populate the galactic halo, then they will be
captured in the Sun and in the Earth \press, annihilate
therein, and produce high-energy neutrinos that could be
detected in underground detectors \neutrinos\marcneutrinos.
Neutralinos from the galactic halo are accreted onto the Sun and
Earth and their numbers therein are depleted by annihilation.
Typically, the two processes come into equilibrium on a time
scale much shorter than the age of the solar system, in which
case the rate for neutralino annihilation is equal to the
capture rate divided by 2, $C/2$.  Then, the differential flux of
energetic neutrinos of type $i$ (e.g. $i=\nu_\mu,\bar\nu_\mu$,
etc.) from neutralino annihilation in the Sun or Earth is
\eqn\simpleflux{
  \left({d\phi\over dE}\right)_i = {C \over 4\pi R^2}
  \sum_F B_F \left({dN\over dE}\right)_{Fi},}
where $R$ is the distance from the detector to the center of the
Sun or Earth, the sum on $F$ is over all annihilation channels,
$B_F$ is the branching ratio for annihilation into channel $F$,
and $(dN/dE)_{Fi}$ is the differential energy spectrum of
neutralino type $i$ at the detector expected from injection of
the particles in channel $F$ at the core of the Sun or Earth
\ritz.  Calculation of these spectra can be quite complicated as
it involves hadronization of the annihilation products;
furthermore, if annihilation takes place in the Sun, interaction
of the annihilation products with the solar medium as well as
interactions of the neutrinos as they propagate through the Sun
must be taken into account.

In the rest of our discussion, we will focus on the neutrino
signal from neutralino annihilation in the Earth.  If the
neutralino is lighter than the top quark, then the neutrino
signal from the Earth should be greater than or equal to that
{}from the Sun (unless the neutralino has only axial interactions
with nuclei in which case it is captured in the Sun but not in
the Earth).  Also, calculation of neutrino spectra from
annihilation in the Earth is much simpler than the calculation
of spectra from the Sun.  It should be kept in mind, however,
that our conclusions will also apply to neutrino rates from the
Sun.

The most promising method of detection of energetic neutrinos is
observation of upward moving muons induced by neutrino
interactions in the rock below the detector.  Given the fluxes
$(d\phi / dE)_i$, the rate for neutrino-induced upward-moving
muons may be written simply as
\eqn\mastereqn{
  \Gamma_{\rm detector}=\sum_i D_i \int\,\left({d\phi\over
  dE}\right)_i E^2 dE,}
where the sum is over $\nu_\mu$, which produce muons, and $\bar
\nu_\mu$, which produce antimuons.  The rate is proportional to
some constant $D_i$, times the second moment of the neutrino
energy spectrum. This is because the cross section for
a neutrino to produce a muon is proportional to the neutrino energy,
and the range of the muon is proportional to its energy, giving an
overall dependence on the square of the impinging neutrino energy.

If neutralino annihilation takes place in the Earth, then
interactions of the annihilation products and
neutrinos with the Earth can be neglected; thus, the
neutrino-energy spectrum from a given annihilation channel $F$
can be determined in a straightforward way from the results of
Monte Carlo calculations \ritz.  Furthermore, in this case, the
energy spectra of neutrinos and antineutrinos are the same, so
the rate for neutrino-induced upward-moving muons from
neutralino annihilation in the Earth may be written
\eqn\earthrate{
  \Gamma_{\rm detector} = 3.9\times10^{-20}\,\left({C \over {\rm
  sec}^{-1}} \right) \left({\Mx \over {\rm GeV}} \right)^2 \sum_F B_F
  \VEV{Nz^2}_F\, {\rm m}^{-2}\,{\rm yr}^{-1}.}

The quantity $\VEV{Nz^2}_F$ is the second moment of the energy spectrum
of neutrinos produced from final state $F$, divided by the neutralino mass
squared.  Final-state electrons, muons, and $u$, $d$, and $s$
quarks will not produce energetic neutrinos.  The weak decays of
$\tau$ leptons and $c$ and $b$ quarks produce energetic
neutrinos, and expressions for $\VEV{Nz^2}$ for these final
states have been given by Ritz and Seckel \ritz.  For neutralino
masses large compared with the light-fermion masses,
$\VEV{Nz^2}_{\bar b b}=0.0195$, $\VEV{Nz^2}_{\bar c c}=0.0084$,
and $\VEV{Nz^2}_{\bar\tau\tau}=0.0682$; thus
the $\bar\tau \tau$ final state gives the strongest
neutrino signal of the fermionic final states.
If the neutrino annihilates into gauge-boson
pairs, then energetic neutrinos are produced directly by the
decays of the gauge bosons, and $\VEV{Nz^2}_{W^+
W^-}=0.035[1-(m_W^2/4\Mx^2)]$ and $\VEV{ Nz^2}_{ZZ}=0.045[1 -
(m_Z^2/4\Mx^2)]$ \marcneutrinos. Expressions for $\VEV{Nz^2}$
{}from final states with Higgs bosons may also be given
\neutrinos.  For annihilation in the Earth, the second moment of
the neutrino spectrum from a Higgs boson $B$ is roughly
$\VEV{Nz^2}_B \simeq \sum_f \VEV{Nz^2}_f \Gamma_f^B/2$, where the sum on
$f$ is over all the decay channels of the $B$ boson,
and $\Gamma_f^B$ is the branching ratio for $B$ decay into final
state $f$.  The branching ratios are given in, e.g.,
Ref.~\higgs.  Leading-order QCD corrections to these results may
be included by using running quark masses instead of the
tree-level quark masses.  Although we will not need it here,
$\VEV{Nz^2}$ for the top-quark final state can be obtained by
noting that the top will decay predominantly into a $b$ quark
and a $W$ boson and using the results for $\VEV{Nz^2}$ for these
final states.

Given Eq.~\earthrate\ and the estimates of $\VEV{Nz^2}$ above,
we can see qualitatively the effect of including the $gg$ and $g
\bar q q$
annihilation channels, as well as the effect of running-quark
masses on rates for energetic-neutrino events from neutralino
annihilation in the Earth.  The gluons will produce essentially
no energetic neutrinos.  Energetic neutrinos will come from weak
decays of heavy quarks, and only a small fraction of
the gluon energy goes into heavy quarks.  Thus, if the cross
section for annihilation into gluons is appreciable, the
branching ratios $B_F$ into the annihilation channels that {\it
do} produce energetic neutrinos are decreased, and the event
rate, given by Eq.~\earthrate, is decreased accordingly.

The effect of annihilation into $g \bar q q \ (q=b,c)$ is
similar, though not as severe, as
the effect of annihilation into gluons.  Again, the emitted
gluon will not produce any energetic neutrinos.  The quark and
antiquark will produce neutrinos, but their energies will be
decreased leading to a smaller event rate.
In the heavy-squark limit,
$r_{\tilde q}= \Msq^2/\Mx^2 \ra \infty$, the
gluon carries away 1/3 the available energy, which suggests that
the rate for
energetic-neutrino events from the 3-body final state containing
$c$ or $b$ quarks would be
roughly 4/9 what it would be if the neutralino
annihilated into $\bar q q$.  As the ratio $r_{\tilde q}$ is
decreased towards unity, the gluon carries away a larger
fraction of the available energy, thereby weakening the
neutrino signal. Moreover, more than 50\% of the total $q \bar q g$
contribution comes from light ($u,d,s$) quarks.
Therefore, the energetic-neutrino flux from
the 3-body final state is generally small: If $r_{\tilde q}\sim 1$, then the
branching ratio may be significant, but the neutrino signal is
small; if $r_{\tilde q}$ is large, the branching ratio into the 3-body
final state is suppressed.
We have not done the calculation of the neutrino spectrum from this
final state more
precisely than indicated here. A more precise calculation would simply
involve a convolution of the neutrino energy spectra from
quark-antiquark pairs with the differential cross section in
Eq.~\qqgcv.

\ifIncludeEPSF
{\topinsert
        \epsfxsize=4in
        \centerline{\vbox{\epsfbox{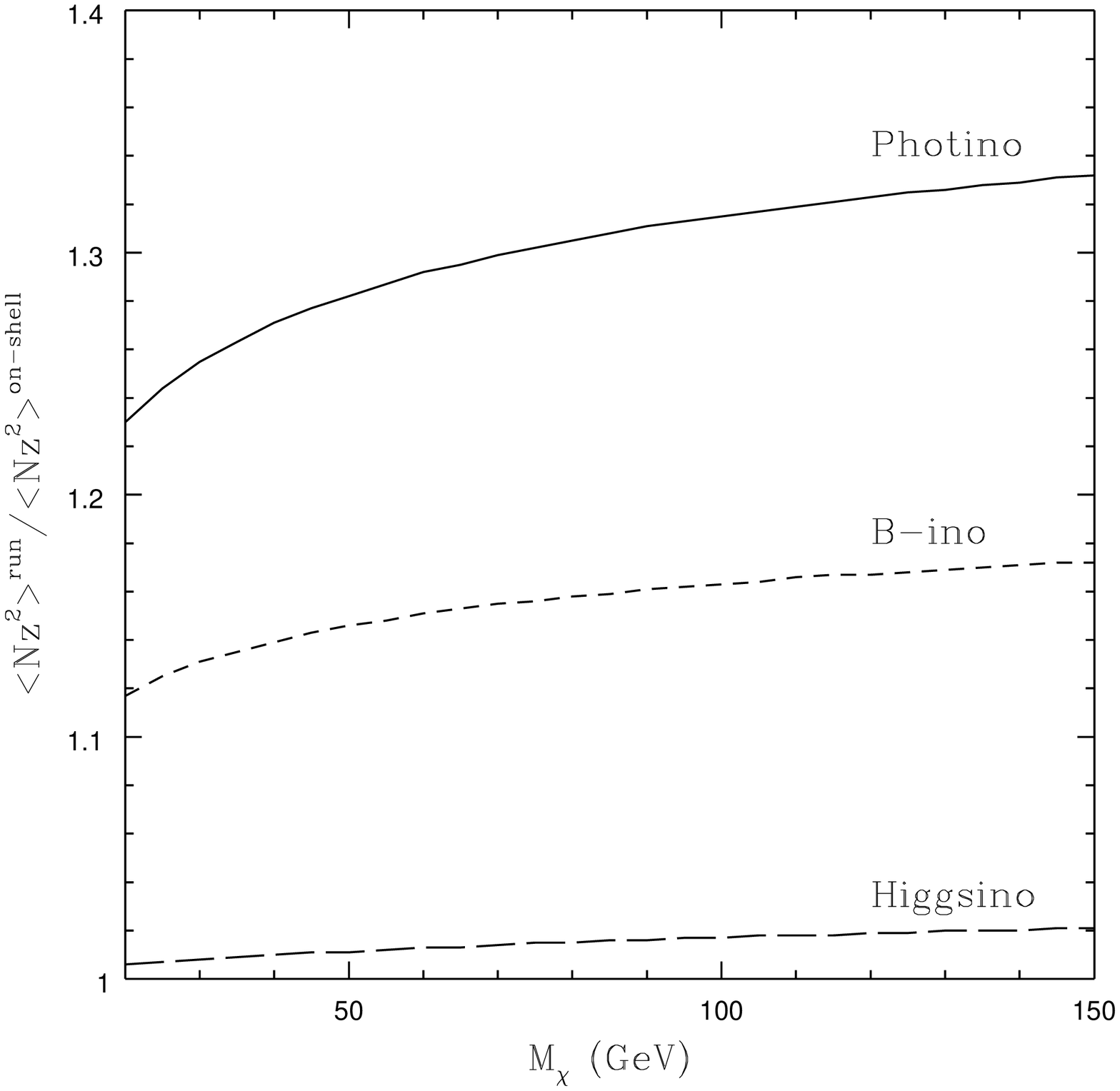}}}
        \vskip 0.2in
        \iFig\ratioplot{Effect of QCD corrections to the
                tree-level cross sections for $\neut\neut\ra
                \bar q q$ on energetic-neutrino event rates.  We
		illustrate the results for the cases of a pure
		photino, pure $B$-ino, and pure higgsino.}
        \epsfxsize=0pt
        \epsfysize=0pt
        \vskip .5cm
\endinsert
}
\else
        \nFig\ratioplot{Effect of QCD corrections to the
                tree-level cross sections for $\neut\neut\ra
                \bar q q$ on energetic-neutrino event rates.  We
		illustrate the results for the cases of a pure
		photino, pure $B$-ino, and pure higgsino.}

\fi

Although the new annihilation channels tend to decrease the
event rate, leading-order QCD corrections to the tree-level
cross section for neutralino annihilation into $b$ and $c$ quarks have
the opposite effect.  QCD corrections decrease the cross
sections for annihilation into $b$ and $c$ quarks, and therefore
increase the branching ratio into $\bar\tau\tau$ leptons.
$\tau$ leptons provide a stronger signal, so the neutrino
event rate is increased.  In \ratioplot, we illustrate this
effect by plotting the ratio of the neutrino event rate obtained
using running quark masses divided by that obtained using on-shell
quark masses.  We illustrate the results for the cases of a pure
photino, pure $B$-ino, and a pure higgsino.
In all three cases, we have assumed annihilation
occurs only into light fermions.  The effect is largest for the
photino.  This is because the cross sections for annihilation of
photinos into $\tau$ leptons is comparable to that into quarks.
For $B$-inos, the branching ratio to $\tau$ leptons is
larger, and for higgsinos, the branching ratio to the $b$ quark
is larger.  Recall that the energetic-neutrino flux from a
given annihilation channel depends on the branching ratio into
that channel, and {\it not} on the total annihilation cross
section.

\ifIncludeEPSF
{\topinsert
        \epsfxsize=4.5in
        \centerline{\vbox{\epsfbox{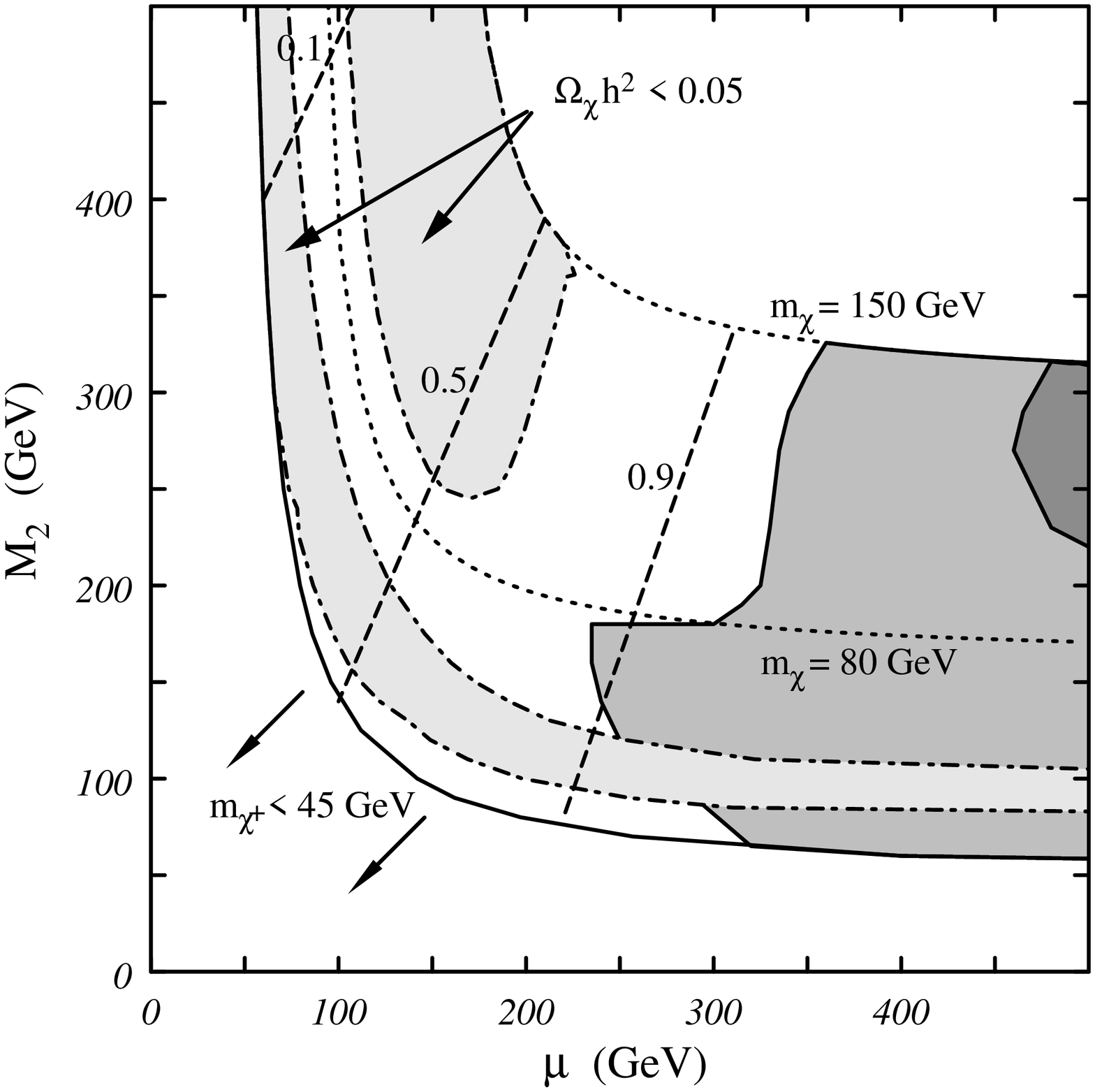}}}
        \vskip 0.2in
        \iFig\nzplot{Combined effect of new results on
                energetic-neutrino event rates.
                The darkest
		region corresponds to
		$\VEV{Nz^2}_{QCD}/\VEV{Nz^2}_{tree} < 0.5$, the next
		lightest to $\VEV{Nz^2}_{QCD}/\VEV{Nz^2}_{tree} < 0.9$,
		and the lightest region corresponds
		to $\Omega_\chi h^2 < 0.05$, as in the previous plots. }
        \epsfxsize=0pt
        \epsfysize=0pt
        \vskip .5cm
\endinsert
}
\else
        \nFig\nzplot{Combined effect of new results on
                energetic-neutrino event rates.
                The darkest
		region corresponds to
		$\VEV{Nz^2}_{QCD}/\VEV{Nz^2}_{tree} < 0.5$, the next
		lightest to $\VEV{Nz^2}_{QCD}/\VEV{Nz^2}_{tree} < 0.9$,
		and the lightest region corresponds
		to $\Omega_\chi h^2 < 0.05$, as in the previous plots.
		}
\fi

In \nzplot, we illustrate the combined effect of including the
$gg$ and $g\bar q q$ final states, as well as the QCD corrected
cross sections for annihilation into $\bar q q$ pairs, on the
rate for energetic-neutrino events.  We consider the ratio of
the event rate predicted using our new results to that which
would be predicted ignoring the new annihilation channels and the
QCD corrections to the light-fermion annihilation cross section,
$\VEV{Nz^2}_{QCD}/\VEV{Nz^2}_{tree}$.  In the most heavily
shaded regions (where the neutralino is primarily gaugino),
the ratio is less than 0.5.  In the more lightly shaded regions
the ratio is less than 0.9.  In the rest of
the graph, the ratio is greater than 0.9, but it is nowhere
greater than 1.1.  Again, the most lightly shaded regions are
those where the cosmological abundance is too small to account
for the dark matter in galactic halos.  The SUSY parameters used
here are the same as those used in \ggplot\ and in \qqgplot:
$m_A=500$ GeV, $m_t=150$ GeV, $\tan\beta=2$, and $\Msq=200$.

We have included all
annihilation channels in this graph.  For our numerical work, we
assumed that the $g\bar q q$ final state produced no energetic
neutrinos.  This underestimates the true flux, but we
are confident that, if the calculation were performed more precisely,
the results in \nzplot\ would not be altered.  We believe so
because our estimate of the neutrino yield from the 3-body final
state is small, and because
\ggplot\ and \qqgplot\ suggest the new annihilation
channels are important only when the 3-body final state is
subdominant.  As was the case for \ggplot, \nzplot\ will change
little qualitatively if the squark mass is changed.

The graph illustrates that our
results are most important for gauginos, as discussed in the
previous Section. Also, note that our results are important in
large regions of parameter space where the neutralino makes a
good dark-matter candidate.  Again, we should point out that we
have explored only a restricted region of parameter space.  The
effect of our results on energetic-neutrino fluxes may be larger
or smaller depending on the specific SUSY parameters used.

\newsec{Summary and Discussion}

We have calculated the cross sections for annihilation of
neutralinos into the two-gluon and gluon-quark-antiquark final
states in the nonrelativistic limit.  We have also calculated
QCD corrections to the tree-level matrix elements for
annihilation of neutralinos into quark-antiquark pairs.  These
new results should have little impact on existing neutralino
cosmological-abundance calculations, although they will be
significant for event rates for indirect-detection schemes.

Since neutralino annihilation into light quarks and leptons is
helicity suppressed in the nonrelativistic limit, the rate for
annihilation into gluons, although suppressed by $\alpha_s^2$,
may be comparable to or greater than that for annihilation into
light quarks and leptons.  Very few energetic neutrinos are
produced by hadronization of gluons in the Sun and in the Earth,
so annihilation into gluons tends to decrease the rate for
energetic-neutrino events.  The new annihilation channels should
have a significant effect if the neutralino annihilates
predominantly into light fermions.  If the neutralino is heavier
than the top quark, or if it is primarily higgsino and heavier
than the $W$ boson, then it will annihilate predominantly into
top quarks or gauge bosons, respectively, and the new channels
should have little effect.

QCD corrections decrease the rate for annihilation into light
quarks.  The flux of energetic neutrinos depends on the
branching ratios into the various final states, and not on the
total annihilation cross section.  In addition, the neutrino
signal from $\bar\tau\tau$ pairs is stronger than that from
light quarks.  Therefore, since the branching ratio into
$\bar\tau\tau$ pairs is increased, while the branching ratios
into light quarks are decreased, QCD corrections to the process
$\neut\neut \ra \bar q q$ increase the energetic-neutrino flux.

The combined effect of the new annihilation channels and QCD
corrections is greater than 10\% in large regions of parameter
space.  If the neutralino is primarily gaugino, the rate for
annihilation into gluons may be greater than the rate for
annihilation into light fermions, and the flux of energetic
neutrinos will be decreased dramatically.  In some regions of
parameter space, the two effects yield a slight increase in the
neutrino flux, although this increase is no greater than about
10\% in any regions of parameter space we explored.

We have not performed a detailed calculation of the flux of
energetic neutrinos that come from the $g\bar q q$ final
state.  If the squark is only slightly heavier than the
neutralino, then the gluon carries away most of the available
energy, and the energetic-neutrino yield should be small.
On the other hand, if the squark mass is large, the branching
ratio for annihilation into the 3-body final state becomes
negligible.

In conclusion, the $gg$ and $g\bar q q$ final states, as well as
QCD corrections to the tree-level amplitudes for annihilation
into $\bar q q$ pairs, are appreciable for many regions of
parameter space.  These new results should therefore be included
in analyses that constrain SUSY dark-matter candidates from
limits on fluxes of energetic neutrinos from the Sun and Earth.

The new results will also have implications for cosmic-ray
antiproton searches \antip.  If neutralinos populate the
galactic halo, they will annihilate and produce low-energy
antiprotons.  Cosmic-ray antiprotons are
produced in standard models of cosmic-ray propagation by
spallation of cosmic rays in the interstellar medium, and
therefore provide a background to the signal from WIMPs.
However, this background decreases dramatically for
cosmic-ray antiproton energies less than about 1 GeV.
There are many astrophysical uncertainties associated with the
predicted fluxes of cosmic-ray antiprotons from WIMP annihilation, so
nonobservation of such cosmic rays cannot be used to eliminate
dark-matter candidates; on the other hand, under certain
reasonable assumptions about the relevant astrophysics and
particle physics, the predicted flux of low-energy cosmic-ray
antiprotons will be large enough to be distinguished from
background.  Observational upper limits to the
cosmic-ray antiproton flux are currently about an order of magnitude
larger than the background expected \pbar.  By performing
similar cosmic-ray experiments at the South Pole, the
sensitivity of these experiments can be improved by several
orders of magnitude \tarle, so a cosmic-ray-antiproton signature
for neutralino dark matter in the halo should be observable,
should such a signature exist.

Our results will have an effect on the cosmic-ray antiproton
flux from neutralino annihilation in the halo with
magnitude similar to that on rates for energetic-neutrino events
{}from annihilation in the Sun and Earth; however, the effect is
converse to that discussed above.
QCD corrections decrease the
cross sections for
annihilation into light quarks, so the number of antiprotons
produced from hadronization of light quarks will be decreased.
On the other hand, if annihilation into gluons is appreciable,
then the antiproton flux will be increased because hadronization
of gluons will produce antiprotons (although Monte Carlo results
will be needed to determine the precise contribution).  The enhancement
will be most significant if the neutralino is primarily gaugino.
We should also point out that if the new
annihilation channels decrease the neutrino rate, then they will
increase the cosmic-ray antiproton flux.  Therefore, cosmic-ray
antiproton searches provide an excellent complement to
energetic-neutrino searches.

\newsec{Acknowledgments}
We thank G.~Giudice and K.~Griest for very useful
discussions.  M.K. was supported by the Texas National Research
Laboratory Commission, and by the U.S. Department of Energy under contract
DE-FG02-90ER40542.  G.J. was supported by the U.S. D.O.E. under contract
DEFG02-90-ER 40560. M.D. was supported by a Heisenberg Fellowship from the
Deutsche Forschungsgemeinschaft, Bonn, Germany. M.D. and M.M.N.
acknowledge support from the U.S. Department of Energy under contract
DE-AC-76ER00881, as well as from the Wisconsin Research Committee
with funds granted by the Wisconsin Alumni Research Foundation.

\vfill\eject
\ifIncludeEPSF {}
\else
\listfigs
\fi
\vfill\eject
\listrefs
\bye